\magnification=1200 

\headline{\ifnum\pageno=1 \nopagenumbers 
\else \hss\number \pageno \fi} 
 
\overfullrule=0pt
\footline={\hfil}
\font\boldgreek=cmmib10
\textfont9=\boldgreek
\mathchardef\mypsi="0920

\mathchardef\myphi="091E

\def\lsim{\raise0.3ex\hbox{$<$\kern-0.75em\raise-1.1ex\hbox{$\sim$}}}
\def\gsim{\raise0.3ex\hbox{$>$\kern-0.75em\raise-1.1ex\hbox{$\sim$}}}
\baselineskip=10pt 
\vbox to 1,5truecm{}
\parskip=0.2truecm 
\centerline{\bf STRUCTURE FUNCTIONS OF NUCLEI AT SMALL x} \smallskip
\centerline{\bf AND DIFFRACTION AT HERA}\bigskip 

\bigskip \centerline{by}\medskip 
\centerline{{\bf A. Capella, A. Kaidalov}\footnote{*}{Permanent address : ITEP,
B. Cheremushkinskaya 25, 117259 Moscow, Russia}{\bf, C. Merino}\footnote{**}
{Permanent address : Universidade Santiago de Compostela, Dep. F{\'\i}sica de 
Particulas, E-15706 Santiago de Compostela, Spain}{\bf, D. Pertermann}
\footnote{***}{Permanent address : Univ-GH-Siegen, Phys. Dept., D-57068 Siegen,
Germany} {\bf and J. Tran Thanh Van}}  
\smallskip
 \centerline{Laboratoire de Physique Th\'eorique et Hautes Energies
\footnote{****}{Laboratoire associ\'e au Centre National de la Recherche
Scientifique - URA D0063}}  \centerline{Universit\'e de Paris XI, 
b\^atiment 211, 91405 Orsay cedex, France}  
\bigskip \bigskip \bigskip\baselineskip=20pt 
\noindent 
${\bf Abstract}$ \par 
Gribov theory is applied to investigate the shadowing effects in the structure
functions of nuclei. In this approach these effects are related to the process 
of diffractive dissociation of a virtual photon. A model for this diffractive 
process, which describes well the HERA data, is used to calculate the shadowing
in nuclear structure functions. A reasonable description of the $x$, $Q^2$ and
$A$-dependence of nuclear shadowing is achieved. \par

\vbox to 4 truecm{}

\noindent LPTHE Orsay 97-04 \par 
\noindent February 1997
\vfill \supereject
\noindent {\bf 1. \underbar{Introduction}} \par \vskip 5 truemm
Deep inelastic scattering (DIS) on nuclei gives important information on 
distributions of quarks and gluons in nuclei. The region of small Bjorken $x$ 
is especially interesting because partonic clouds of different nucleons overlap 
as $x \to 0$ and shadowing effects become important. There are experimental 
results in this region, which show that there are strong deviations from an 
$A^1$ behavior in the structure functions [1]. Several theoretical models have 
been proposed to understand these data [1]. The most general approach is based
on the Gribov method [2]. It relates partonic and hadronic descriptions of 
small $x$ phenomena in interactions of real or virtual photons with nuclei. In
this approach the shadowing effects can be expressed in terms of the 
cross-sections for diffraction dissociation of a photon on a nucleon (Fig. 1).
This process has been studied recently in DIS at HERA [3]. The detailed $x$, 
$Q^2$ and $M^2$ ($M$ is the invariant mass of the diffractively produced system)
dependencies observed in these experiments [3] have been well described in the
theoretical model of ref. [4] which is based on Regge factorizations and uses 
as an input available information on diffractive production in hadronic 
interactions. Here we will apply the same model to calculate the structure
functions of nuclei in the small $x$-region. The use of the model, which 
describes well the diffraction dissociation of virtual photons on a nucleon 
target, leads to a strong reduction of the theoretical uncertainty in 
calculations of the structure functions of nuclei in comparison with previous 
calculations [1, 5]. \par \vskip 5 truemm

\noindent {\bf 2. \underbar{The model}} \par \vskip 5 truemm

In the Gribov approach the forward scattering amplitude of a photon with 
virtuality $Q^2$ on a nuclear target can be written as the sum of the 
diagrams shown in Fig. 2. The diagram of Fig. 2a corresponds to the sum of 
interactions with individual nucleons and is propotional to $A^1$. The second 
diagram (2b) contains a double scattering with two target nucleons. It gives a 
negative contribution to the total cross-section, proportional to $A^{4/3}$ 
(for large $A$) and describes the shadowing effects for sea quarks. According 
to reggeon diagram technique [6] and Abramovsky, Gribov, Kancheli (AGK) cutting
rules [7], the contribution of the diagram of Fig. 2b to the total 
$\gamma^*$-$A$ cross-section is related to the diffractive production of 
hadrons by a virtual photon as follows~: 
$$\sigma_A^{(2)} = - 4 \pi \ A(A - 1) \int d^2b \ T_A^2(b)
\int dM^2 \left . {d \sigma_{\gamma^*p}^{\cal D} \over dM^2dt} \right |_{t=0}
F_A(t_{min}) \ \ \ , \eqno(1)$$ \noindent where $T_A(b)$ is the nuclear profile
function, $\rho_A$ is the nucleon density 
($T_A(b) = \int_{- \infty}^{+\infty} dZ
\rho_A(b, Z),\break \noindent \int d^2b \ T_A(b) = 1$) and
$$F_A(t_{min}) = \int dZ \ e^{-iq_ZZ} \rho_A (b, Z)/T_A(b) \ ,
\ t_{min} = - q^2_Z = - m_N^2 \ x^2 
\left ( {Q^2 \over M^2 + Q^2} \right )^{- 2} \ \ \
.$$ 

\noindent Note that
$F_A(t_{min})$ is equal to unity as $x \to 0$ and decreases fast as $x$ 
increases to $x_{cr} \sim {1 \over m_N R_A}$, due to a lack of coherence for 
$x > x_{cr}$. \par

Thus the second order rescattering term can be calculated if the differential
cross-section for diffractive production by a virtual photon is known. \par

Higher order rescatterings are model dependent, but in the region of 
$x \gsim 10^{-3}$,
where experimental data exist, their contribution is rather small 
($\sim$ several $\%$) for existing nuclei. We use the following unitary 
expression for the total $\gamma^*$-$A$ cross-section
$$\sigma_{\gamma^*A} = \sigma_{\gamma^*N} \int d^2b {A \ T_A(b) 
\over 1 + (A - 1) f(x,Q^2) T_A(b)} \eqno(2)$$
\noindent where
$$f(x, Q^2) = 4 \pi \int dM^2 \left . 
{d\sigma_{\gamma^*p}^{\cal D} \over dM^2dt} \right
|_{t=0} F_A(t_{min})/\sigma_{\gamma^*N} \ \ \ . $$
\noindent This expression is valid in the generalized Schwimmer model [8, 9] 
and corresponds to the summation of fan diagrams with triple Pomeron 
interaction. We have checked that the results for eikonal-type summation of 
higher order rescatterings are very similar to the one obtained with (2). \par

Thus we obtain for the ratio $R_A = F_{2A}/F_{2N}$, in the region of small $x$ 
$${F_{2A} \over F_{2N}} = \int d^2b {A \ T_A(b) 
\over 1 + (A - 1) f(x, Q^2) T_A(b)} \ \
\ . \eqno(3)$$
\noindent The deviation of this ratio from $A^1 = A \int d^2b T_A(b)$ is due to 
the second term in the denominator of the integrand in eq. (3). Thus, knowing 
the differential cross-section for diffraction dissociation on a nucleon and 
the structure function of a nucleon ($\sigma_{\gamma^*N})$, one can predict the
$A$ (and $x$, $Q^2$) dependence of structure functions of nuclei. Eq. (3) can 
only be used in the region $x < 10^{-1}$ where the sea quarks component 
dominates, because at larger values of $x$ shadowing of valence quarks (which 
in general is not described by eq. (3)) becomes important [10, 11]. The effects
which lead to antishadowing are also important in the region of 
$x \sim 0.1$. \par

In refs [4] we described the diffractive contribution to DIS in terms of 
Pomeron exchange
$$F_2^{\cal D}(x, Q^2, x_P, t) = {(g_{pp}^P(t) )^2 \over 16 \pi} 
x_P^{1-2\alpha_P(t)} F_P(\beta, Q^2, t) \eqno(4)$$
\noindent where $g_{pp}^P(t)$ is the Pomeron-proton coupling $(g_{pp}^P(t) =
g_{pp}^P(0) \exp (Ct)$ with $(g_{pp}^P(0))^2 =$ 23~mb and 
$C = 2.2$ GeV$^{-2}$), $\alpha_P(t)$ is the Pomeron trajectory 
($\alpha_P(0)$ = 1.13, $\alpha '_P(0) = 0.25$ GeV$^{-2}$) and 
$F_P(\beta , Q^2, t)$ is the Pomeron structure function. The variable
$\beta = {Q^2 \over M^2 + Q^2} = {x \over x_P}$ plays the same role for the 
Pomeron as the Bjorken variable $x$ for the proton. At large $Q^2$, $F_P$ can 
be expressed in terms of the quark distribution in the Pomeron
$$F_P(\beta , Q^2,t) = \sum_i e_i^2 \beta \left [ q_i^P(\beta , Q^2, t) +
\bar{q}_i^P(\beta , Q^2, t) \right ] \ \ \ .  \eqno(5)$$
\noindent In refs. [4] we determined $F_P(\beta , Q^2,t)$ using 
Regge-factorization for small values of $\beta$ and a plausible assumption on 
the $\beta \to 1$ behavior.
This function was then used as an initial condition for QCD evolution of 
partons in the Pomeron. The results of the QCD-evolution crucially depend on 
the form of the gluon distribution in the Pomeron. Experimental results for 
$F_2^{\cal D}$ can be understood only if the distribution of gluons in the 
Pomeron is rather hard and the gluons carry the main part of the Pomeron 
momentum [4, 12-14]. \par

The model described above gives a reasonable description of the process of 
diffractive production in DIS and can be used to compute the function 
$f(x, Q^2)$, which determine the shadowing of nuclear structure function via 
(3). This function can be written in terms of the ratio $F_P/F_{2N}$~: 
$$f(x, Q^2) = \int {d\beta \over
4 \beta} \left ( g_{pp}^P(0) \right )^2 \left 
( {1 \over x_P} \right )^{2\Delta}
{F_P(\beta , Q^2) \over F_{2N} (\beta , Q^2)} F_A(t_{min}) \eqno(6)$$
\noindent where the integral limits are $x/x_{0P}$ with $x_{0P} = 0.1$ and
$Q^2/(m_{\rho}^2 + Q^2)$. For the nucleon structure function $F_{2N}$ we use 
the results of ref. [15] and for $F_P(\beta , Q^2)$ the expression and 
parameters of ref. [4b]. In the numerical calculations we use a standard 
Woods-Saxon profile for $A > 20$ and a gaussian for $A < 20$ with a rms radius
given by $R_A = 0.82 A^{1/3} + 0.58$ fm [16]. For deuteron we have also used a
gaussian with an artificially high radius in order to reproduce the small 
amount of shadowing present in this case. The results of our calculations are 
shown in Figs. 3-7. \par \vskip 5 truemm

\noindent {\bf 3. \underbar{Numerical results}} \par \vskip 5 truemm

Comparison of our predictions for the ratio ${2 \over A} F_{2A}/F_2^D$ with
experimental data of NMC [17] is shown in Fig. 3 and for ratios of different 
nuclei in Fig. 4. New data for the ratio $F_2^{S_n}/F_2^C$ [18] are also shown
in Fig. 4. It is important to note that experimental points in Figs. 3, 4 for 
different $x$ correspond to different values of $<Q^2>$ [17] [18]. This 
correlation has been taken into account in our calculations. The agreement 
between theoretical predictions and experimental data is rather good taking 
into account that there are no free parameter in our calculations. However, one
should notice that for large nuclei ($S_n$), the onset of the shadowing seems 
to move to smaller values of $x$ as compared to light ones. This trend is also
present in the $Pb$ data of ref. [19] and is not reproduced in our model 
(despite the decrease of $x_{crit}$ with increasing $A$). This probably 
indicates that antishadowing (not present in our model) extends to smaller $x$
values for heavier nuclei. \par

Our predictions for the ratio ${1 \over A} {F_{2A} \over F_{2N}}$ in the region
of very small $x$ are shown in Figs. 5 for fixed values of $Q^2$. They can be
confronted to experiment if nuclei at HERA will be available. Note that our 
results are more reliable for small values of $x$ ($x < 10^{-2}$) where the 
effects of both valence quark shadowing and antishadowing are negligeable. The
curves for shadowing effects in the gluon distribution of nuclei ${1 \over A} 
{g^A(x, Q^2) \over g^N(x,Q^2)}$ are shown in Figs. 6. They look similar to the
shadowing for the quark case. However the absolute magnitude of the shadowing 
is smaller in the gluon case contrary to expectations of some theoretical 
models [1] but in agreement with [20]. (Note that these results are sensitive 
to the gluonic distribution in the Pomeron, which is poorly known at present).
These predictions can be tested in experimental studies of $J/\psi$ and 
$\Upsilon$-production on nuclear targets at RHIC and LHC. \par

Finally we want to discuss in more detail the $Q^2$-dependence of the 
shadowing. Recent NMC data [18] for the ratio of $F_2^{S_n}/F_2^C$ are shown in
Figs. 7 as functions of $Q^2$ for fixed values of $x$ in the small $x$ region.
The theoretical curves have a weak dependence on $Q^2$ and are in a reasonable
agreement with experiment, although the $Q^2$ dependence seems stronger in the
data especially in the region of small $Q^2$. At larger values of $x$ the data
are practically $Q^2$-independent. These properties should be checked in future
experiments. \par
\vskip 5 truemm

\noindent {\bf 4. \underbar{Conclusions}} \par \vskip 5 truemm

In conclusion, a model based on the Gribov-Glauber theory of nuclear shadowing
and the properties of diffraction in DIS observed at HERA, leads to a fair 
description of experimental data on structure functions of nuclei in the small
$x$ region. Predictions of shadowing effects for quark and gluon distributions
are given. They can be tested in future experiments at HERA and in hadronic 
colliders. \par \vskip 5 truemm

\noindent {\bf Acknowledgements} \par
One of us (A.C.) would like to thank G. Do Dang for discussions. A. K. and 
D. P. wish to thank the LPTHE for hospitality during a period when this work 
was initiated. The work has been partially supported by grant 93-0079 of INTAS.
A. K. also acknowledges support from the grant N$^{\circ}$ 96-02-19184 of RFFI.

\vfill\supereject \centerline{\bf References} \vskip 3 truemm 
\item{[1]} M. Arneodo, Phys. Reports {\bf 240} (1994) 301 (and references 
therein).
\item{[2]} V. N. Gribov, ZhETF {\bf 56} (1969) 892, ibid {\bf 57} (1969) 1306 
[Sov. Phys. JETP {\bf 29} (1969) 483, {\bf 30} (1970) 709]. 
\item{[3]} T. Ahmed et al (H1 collaboration), Phys. Lett. {\bf B348} (1995) 681.
 \item{} M. Derrick et al (Zeus collaboration), Z. Phys. {\bf C68} (1995) 569~;
         Z. Phys. {\bf C70} (1996) 391. 
 \item{[4]} a) A. Capella, A. Kaidalov, C. Merino and J. Tran Thanh Van, Phys.
               Lett. {\bf B343} (1995) 403.  
\item{} b) A. Capella, A. Kaidalov, C. Merino, D. Pertermann and J. Tran Thanh
Van, Phys. Rev. {\bf D53} (1996) 2309. \item{[5]} K. Boreskov, A.
Capella, A. Kaidalov and J. Tran Thanh Van, Phys. Rev. {\bf D47} (1993) 219. 
\item{[6]} V. N. Gribov, ZhETF {\bf 57} (1967) 654 [Sov. Phys. JETP {\bf 26} 
           (1968) 14]. 
\item{[7]} V. A. Abramovsky, V. N. Gribov and O. V. Kancheli, Yad. Fiz. 
           {\bf 18} (1973) 595 [Sov. J. Nucl. Phys. {\bf 18} (1974) 308]. 
\item{[8]} A. Schwimmer, Nucl. Phys. {\bf B94} (1975) 445. 
\item{[9]} K. G. Boreskov et al., Yad. Fiz. {\bf 53} (1991) 569 [Sov. J. Nucl.
           Phys. {\bf 53} (1991) 356].   
\item{[10]} L. L. Frankfurt, M. I. Strikman and S. Liuti, Phys. Rev. Lett. 
            {\bf 65} (1990) 1725. 
\item{[11]} A. B. Kaidalov, C. Rasinariu and U. Sukhatme, UICHEP-TH/96-9. 
\item{[12]} T. Gehrmann and W. J. Stirling, Z. Phys. {\bf C70} (1996) 89.  
\item{[13]} K. Golec-Biernat and J. Kwiecinski, Phys. Lett. {\bf B353} (1995) 
            329. 
\vfill \supereject
\item{[14]} J.
Dainton (H1 collaboration), Proceedings Workshop on Deep Inelastic Scattering 
and QCD, Paris, France, 24-28 April 1995 (ed. J. P. Laporte and Y Sirois). 
\item{[15]} A. Capella, A. Kaidalov, C. Merino, J. Tran Thanh Van, Phys. Lett 
            {\bf B337} (1994) 358.
\item{[16]} M. A. Preston and R. K. Bhoduri, Structure of the Nucleus, 
            Addison-Wesley, New York 1975. \item{[17]} P. Amandruz et al 
            (NMC collaboration), Nucl. Phys. {\bf B441} (1995) 3. 
\item{[18]} N. Arneodo et al (NMC collaboration), Nucl. Phys. {\bf
            B481} (1996) 23.  
\item{[19]} M. R. Adams et al (E665 collaboration), Z. Phys. 
            {\bf C67} (1995) 407. 
\item{[20]} K. J. Eskola, Nucl. Phys. {\bf B400} (1994) 240.

\vfill \supereject
\centerline{\bf Figure Captions}
\vskip 3 truemm
{\parindent= 2 truecm
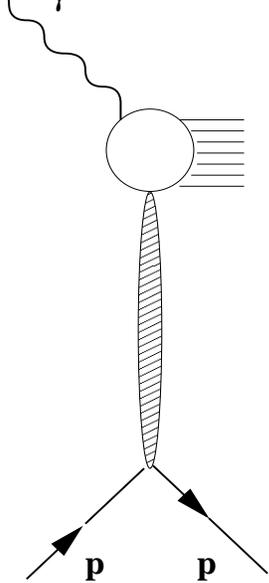
\item{\bf Fig. 1 :} Diffractive dissociation of a virtual photon. The shaded 
     area represents the exchange of a Pomeron. 
\vskip 3 truemm

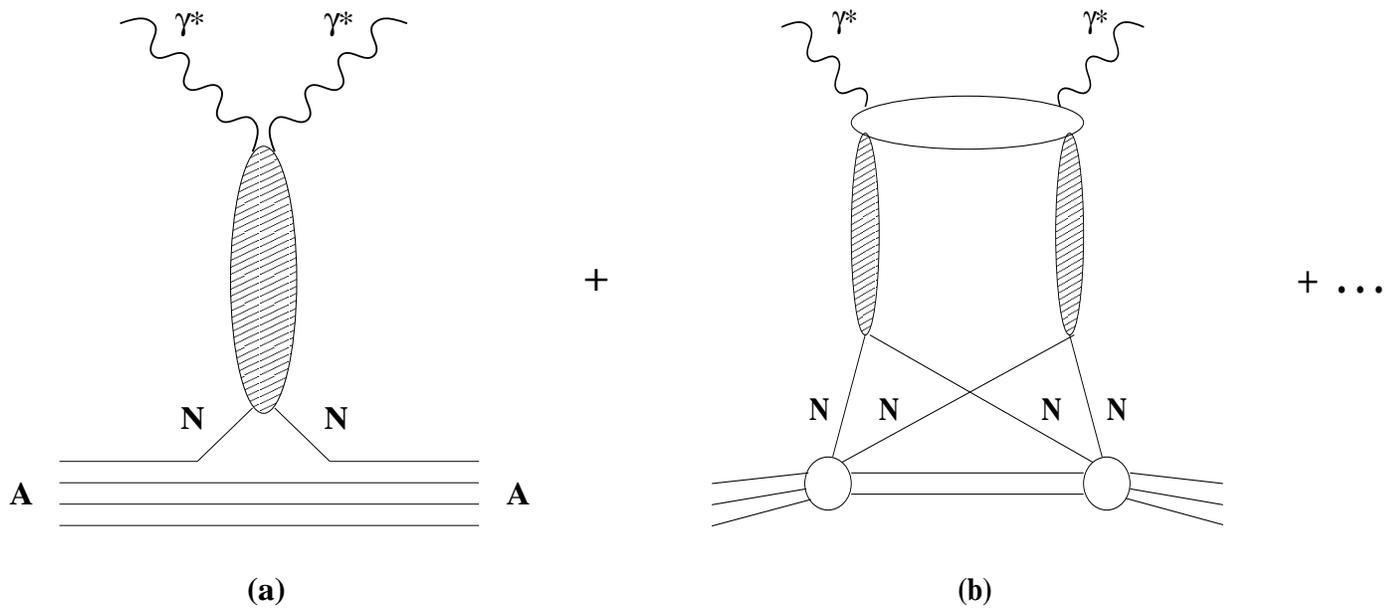
\item{\bf Fig. 2 :} The first two terms (single and double scattering) of the
multiple scattering series for the total $\gamma^*N$ cross-section in the
Gribov-Glauber approach.
\vskip 3 truemm

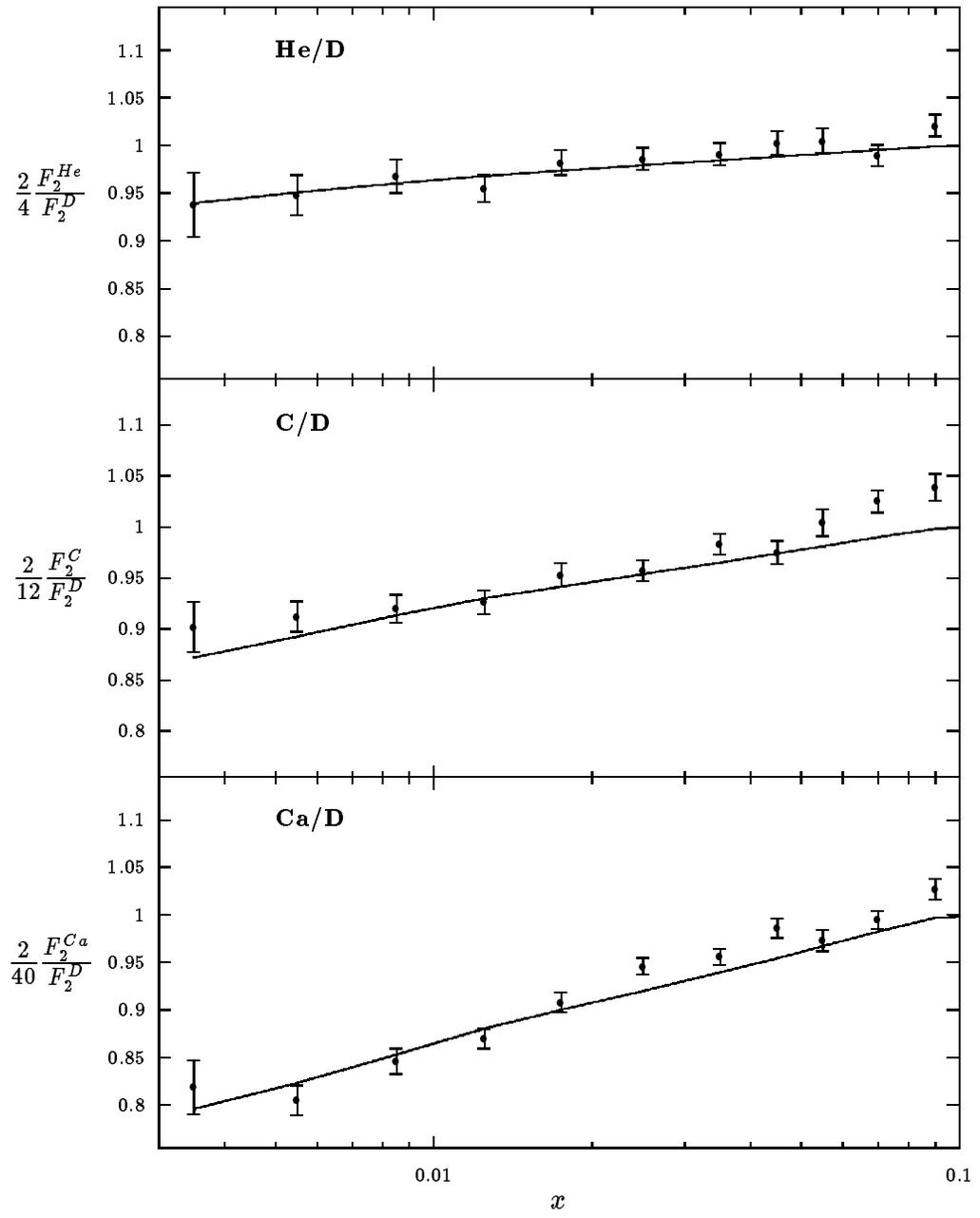
\item{\bf Fig. 3 :} The ratios $(2/A) F_2^A/F_2^D$ computed from eq. (3) for 
     different values of $x$. The experimental points are from ref. [17]. The 
     values of $Q^2$ are different for different $x$-values [17].

\vskip 3 truemm
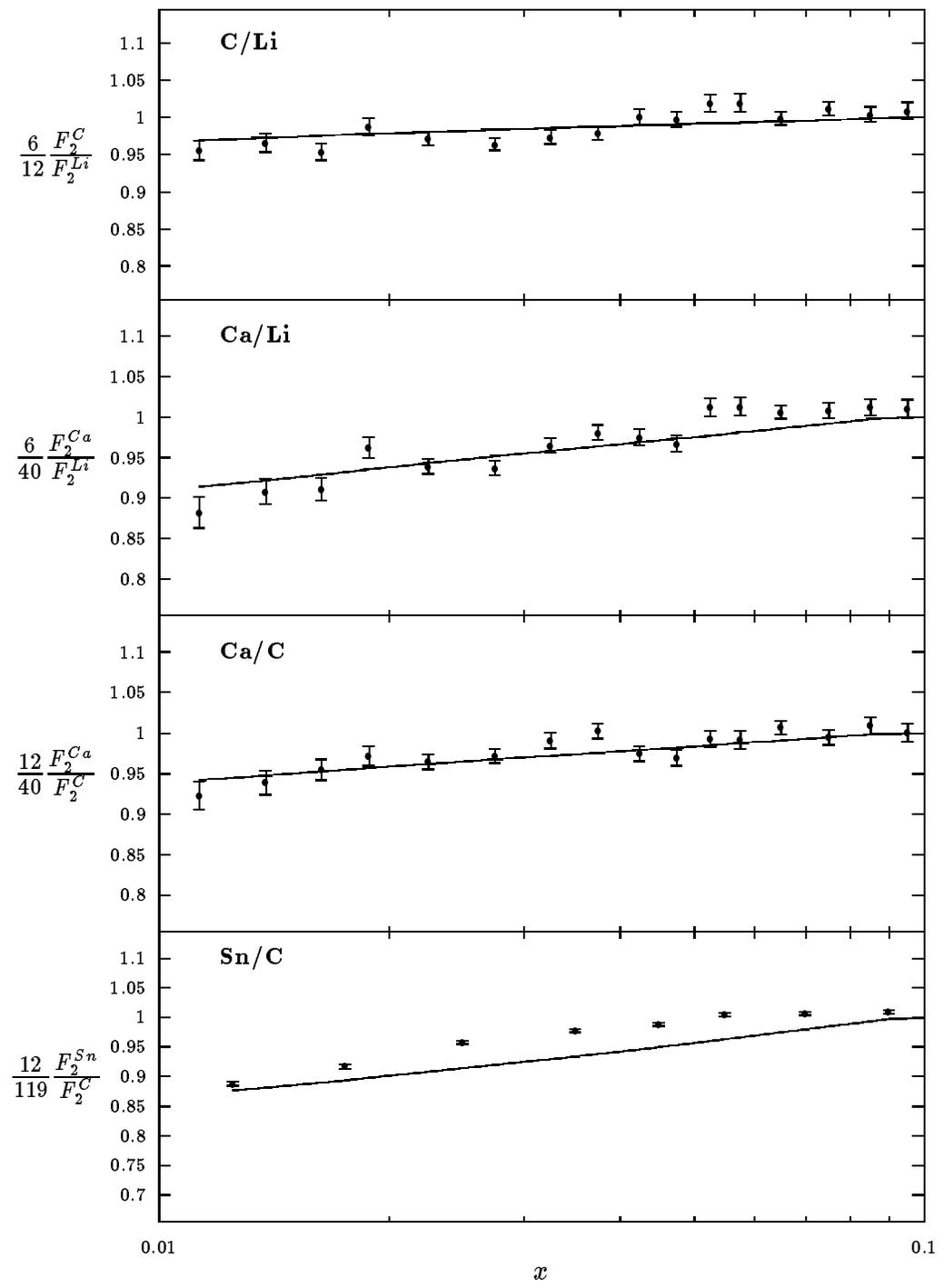
\item{\bf Fig. 4 :} The ratios $(A_1/A_2) F_2^{A_2}/F_2^{A_1}$ computed from 
     eq. (3) for different values of $x$. The experimental points are from 
     refs. [17] and [18]. The values of $Q^2$ are different for different $x$ 
     values [17, 18].

\vskip 3 truemm
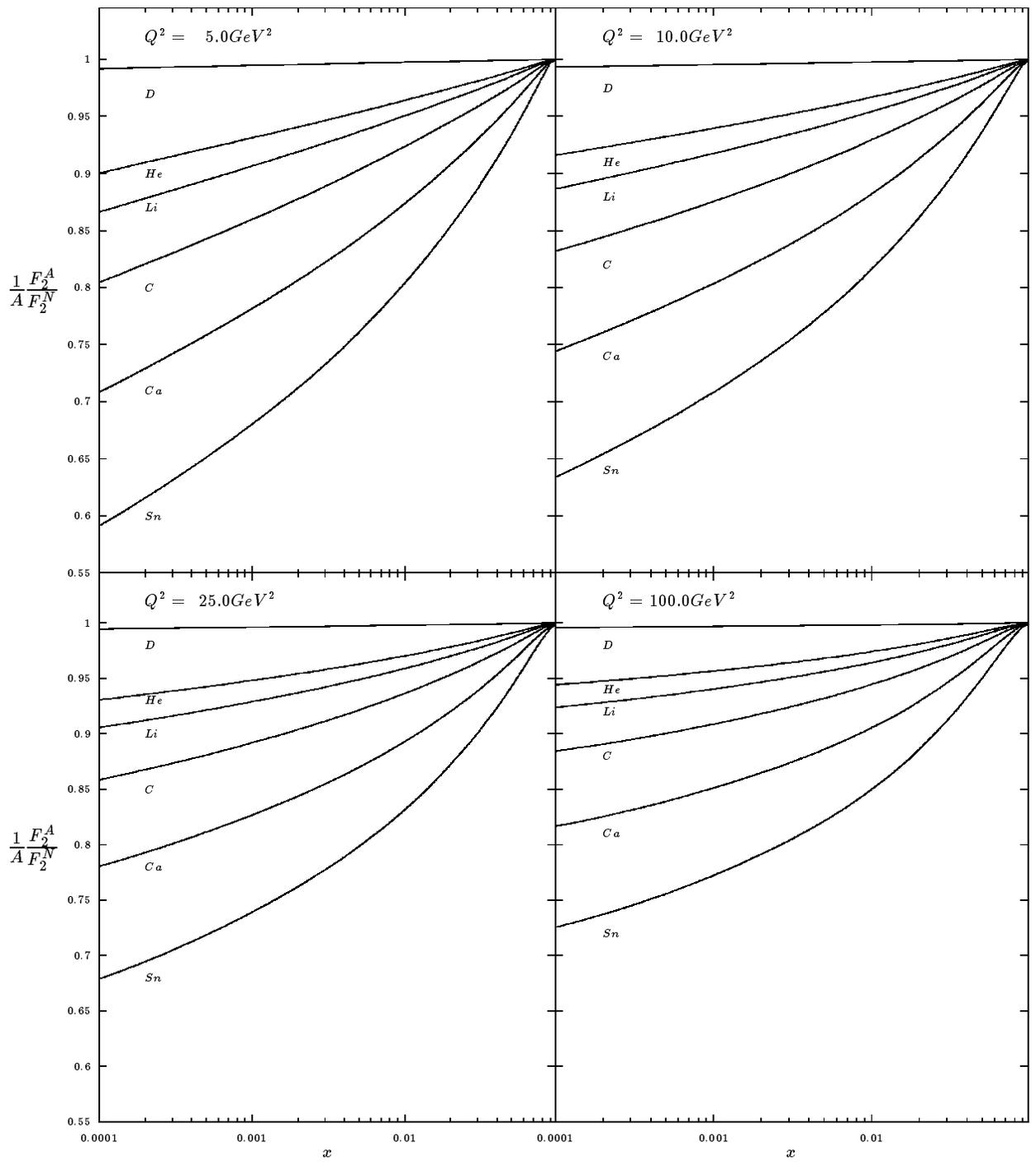
\item{\bf Fig. 5 :} The ratios $(1/A) F_2^{A}/F_2^{N}$ computed from eq. (3) 
     for different values of $x$ in the small $x$ region, at fixed values of 
     $Q^2$.

\vskip 3 truemm
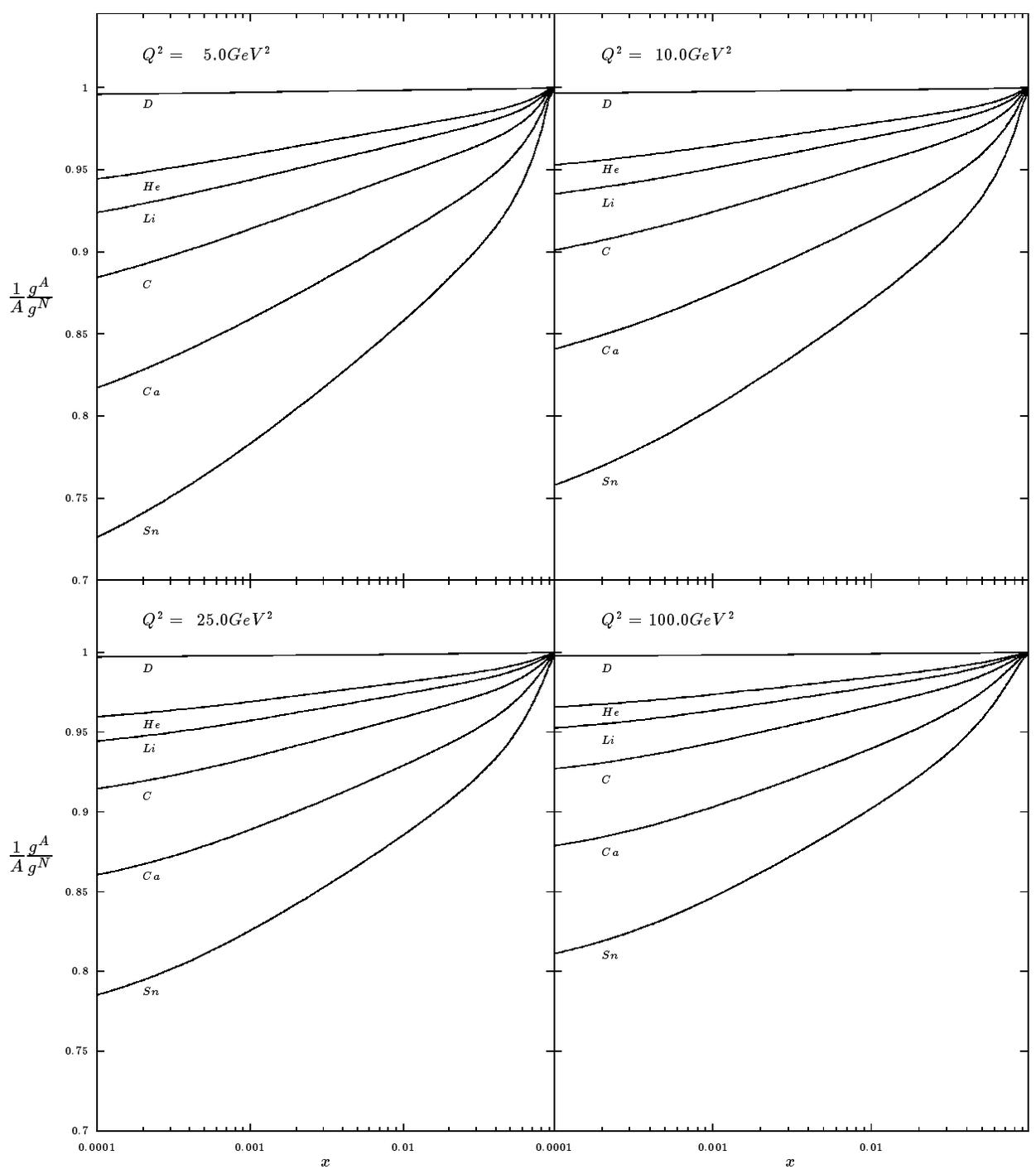
\item{\bf Fig. 6 :} The ratios $(1/A_2) g^A/g^N$ of gluon distribution 
     functions computed for different values of $x$ in the low $x$ region, at 
     fixed values of $Q^2$.

\vskip 3 truemm
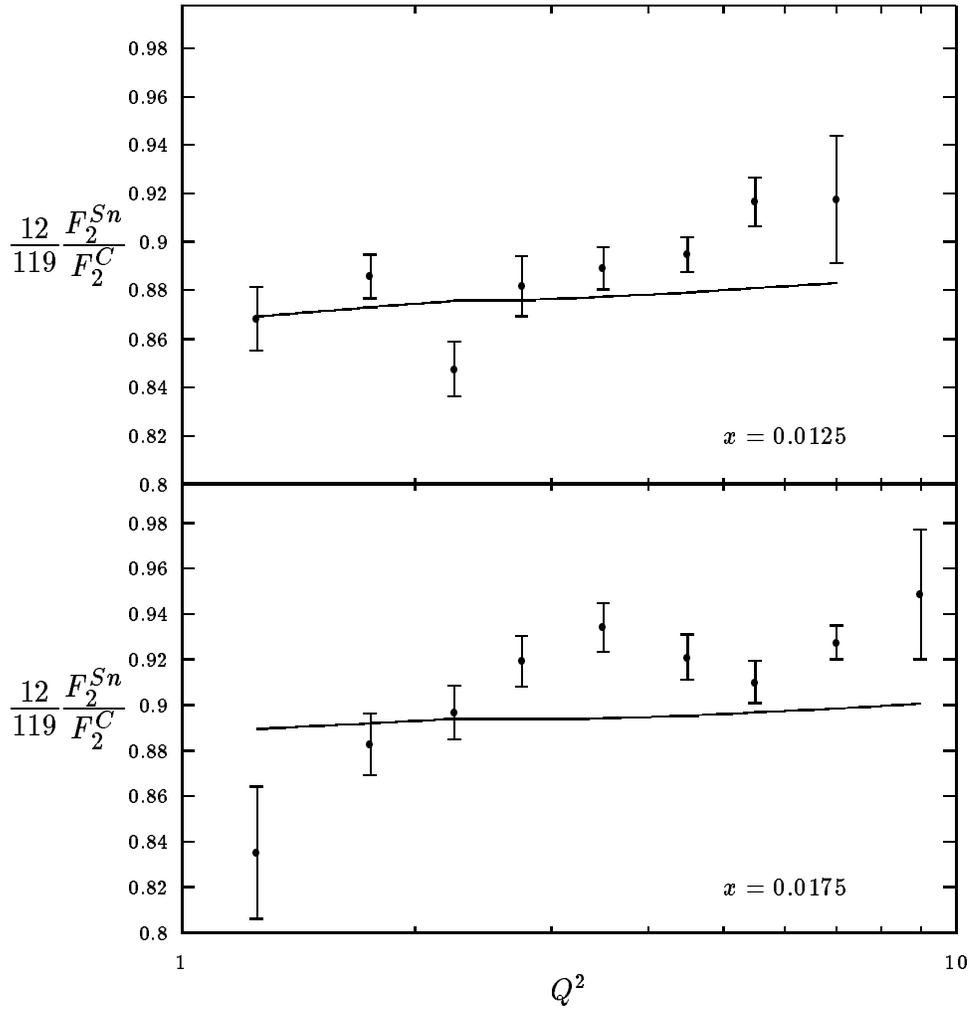
\item{\bf Fig. 7 :} The ratio $(12/119) F_2^{S_n}/F_2^C$ computed from eq. (3) 
     for different values of $Q^2$, at two fixed values of $x$. The data points
     are from ref. [18]. \par
}

\bye